\documentclass[aps,twocolumn,pre,showpacs]{revtex4}
\usepackage{graphics,amssymb,amsmath}

\begin{document}

\title{Comment on ``Ising model on a small world network''}
\author{ H. \surname{Hong}}
\email{hhong@kias.re.kr}
\affiliation{Korea Institute for Advanced Study, Seoul 130-012, Korea}
\author{Beom Jun \surname{Kim}}
\email{kim@tp.umu.se}
\affiliation{Department of Theoretical Physics, 
   Ume{\aa} University, 901 87 Ume{\aa}, Sweden}
\author{M.Y. \surname{Choi}}
\email{mychoi@phya.snu.ac.kr}
\affiliation{Department of Physics, Seoul National University,
Seoul 151-747, Korea}

\begin{abstract}
In the recent study of the Ising model on a small-world network
by A. P\c{e}kalski [Phys. Rev. E {\bf 64}, 057104 (2001)],
a surprisingly small value of the critical exponent $\beta \approx 0.0001$ 
has been obtained for the temperature dependence of the magnetization.
We perform extensive Monte Carlo simulations of the same model
and conclude, via the standard finite-size scaling of various quantities,
that the phase transition in the model 
is of the mean-field nature, in contrast to the work by
A. P\c{e}kalski but in accord with other existing studies.

\end{abstract}

\maketitle

A. P\c{e}kalski in Ref.~\cite{ref:Pekalski} studied the Ising model 
on a small-world network constructed from a ring lattice.  
In the presence of a finite fraction of additional
long-range interactions, the model was observed to undergo a
phase transition at a finite temperature, in agreement
with other related studies~\cite{ref:Gitt,ref:Barrat,ref:XY}.
However, the exponent $\beta$, describing the critical behavior of 
the magnetization in the vicinity of the transition, 
was found to be very small: $\beta \approx 0.0001$, 
in contrast to the previous studies suggesting the mean-field nature of the 
transition~\cite{ref:Gitt,ref:Barrat,ref:XY}. 
The small-world network in Ref.~\onlinecite{ref:Pekalski} was
constructed in a slightly different way 
compared with the original model by Watts and Strogatz~\cite{ref:WS}, 
under the additional constraint that not more than one shortcut is allowed
for each vertex in the network.  
In this comment we present results of extensive Monte Carlo (MC) simulations of
the same model as that in Ref.~\onlinecite{ref:Pekalski}, 
which reveal that the phase transition is 
described by the mean-field exponents, $\alpha = 0$,
$\beta = 1/2$, $\gamma = 1$, and ${\bar \nu} = 2$ (see below for definitions),
and thus conclude that the additional constraint in the 
network construction does not change the universality class of the transition.
   
For simplicity, we consider only the network where every vertex
has one shortcut (or every spin has three couplings), 
which is identical to P\c{e}kalski's network with
the parameter $p=1$ in Ref.~\onlinecite{ref:Pekalski}.
We then perform extensive Monte Carlo simulations 
of the Ising model described by the Hamiltonian:
\begin{equation}
H = -\frac{J}{2} \sum_i \sum_{j \in \Lambda_i} \sigma_i \sigma_j, 
\end{equation}
where $J$ is the coupling strength, 
$\sigma_i (=\pm 1)$ is the Ising spin on vertex $i$, 
and $\Lambda_i$ denotes the neighborhood of vertex $i$, including those vertices
connected to $i$. 
For given network size $N$ 
at temperature $T$ (in units of $J/k_B$), 
we have measured various thermodynamic quantities such as 
Binder's cumulant~\cite{ref:Binder}, the specific heat, and the susceptibility:
\begin{eqnarray}
U_N & = & 1 - \frac{ [ \langle m^4 \rangle ]}{ 3 [\langle m^2 \rangle]^2} 
   \label{eq:Binder} \\
C_v &=& \frac{[\langle H^2 \rangle - \langle H \rangle^2]}{T^2 N} 
   \label{eq:Cv} \\
\chi &=& \frac{1}{N} \sum_{ij} [ \langle \sigma_i \sigma_j \rangle ] 
   \label{eq:chi} 
\end{eqnarray}
with $m \equiv |(1/N)\sum_i \sigma_i|$, 
where $\langle \cdots \rangle$ 
and $[ \cdots ]$ represent the thermal average (taken over 5000 MC steps after
discarding 5000 MC steps for equilibration at each temperature)  
and the average over different network realizations (taken over 400-1200 different
configurations), respectively. 

\begin{figure}
\centering{\resizebox*{!}{11.0cm}{\includegraphics{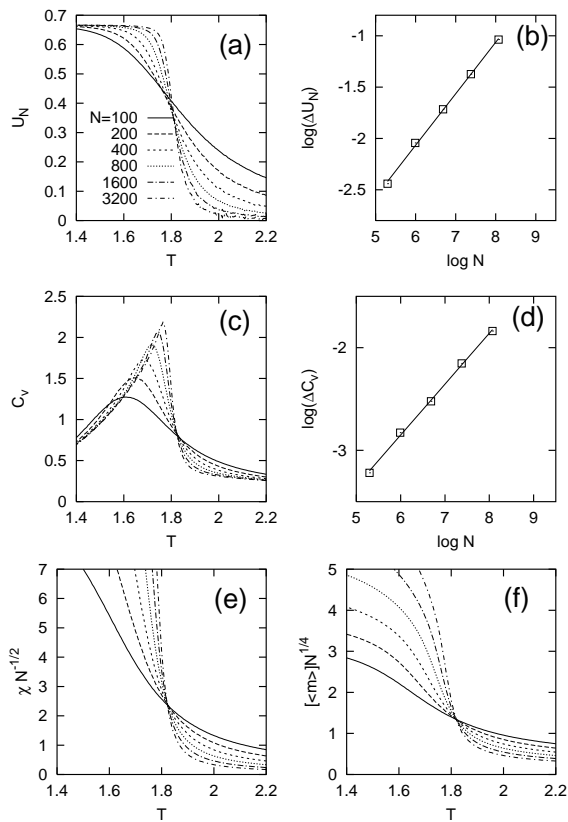}}}
\caption{(a) Binder's cumulant $U_N$ has a unique crossing point at 
$T_c = 1.81(2)$ (in units of $J/k_B$). (b) The critical exponent 
$\bar\nu = 1.99$ is obtained from the least-square fit to the form
in Eq.~(\ref{eq:expand}).
(c) Specific heat $C_v$ (in units of $k_B$) also has a crossing point 
at $T_c = 1.82(2)$, suggesting $\alpha = 0$. (d) From the similar 
expansion of $C_v$ near $T_c$, $\bar\nu = 2.02$ is obtained.
(e) Finite-size scaling of the susceptibility again determines $T_c = 1.82(2)$
with the critical exponent $\gamma = 1$.
(f) Finite-size scaling of the magnetization $[\langle m\rangle]$ 
[see Eq.~(\ref{eq:m})] with $\beta = 1/2$ also leads to the crossing point 
at $T_c = 1.82(1)$. 
In (a), (c), (e), and (f), simulations have been performed with 
the temperature increment $\Delta T = 0.005$ 
whereas the data for $N \geq 200$ have been used
for fitting in (b) and (d).
}
\label{fig:mc}
\end{figure}

Binder's cumulant in Eq.~(\ref{eq:Binder}), plotted for various sizes, 
yields a unique crossing point as a function of the temperature $T$, 
providing a convenient method to determine the critical
temperature $T_c$. 
Figure~\ref{fig:mc}(a) shows that the result
$T_c = 1.81(2)$ is obtained from the crossing point of Binder's cumulant
for $N \geq 800$. 
The critical exponent $\bar\nu$, describing the divergence of the correlation 
volume in such a way that $\xi_V \sim |T - T_c|^{-\bar\nu}$~\cite{ref:XY}, 
can be determined from the expansion of $U_N$ near $T_c$~\cite{ref:HXY}:
\begin{equation} \label{eq:expand}
\Delta U_N \equiv U_N (T_1)-U_N (T_2) \propto N^{1/\bar\nu},
\end{equation}
where $T_1$ and $T_2 \,(>T_1)$ are chosen near $T_c$.  
Figure~\ref{fig:mc}(b) results in
the value ${\bar\nu} \approx 2.0$, which, 
together with the hyperscaling relation $\bar\nu = 2 - \alpha$, 
gives the critical exponent $\alpha \approx 0$ for the specific heat $C_v$. 

With such a mean-field value, we write the finite-size scaling in the form
\begin{equation} \label{eq:Cvscale}
C_v = f\Bigl( (T{-}T_c)N^{1/{\bar\nu}}\Bigr),
\end{equation}
where $f(x)$ is an appropriate scaling function with the scaling variable $x$. 
As shown in Fig.~\ref{fig:mc}(c), the unique crossing point of $C_v$
again yields $T_c = 1.82(2)$, in accord with $T_c$ obtained 
from $U_N$ within numerical errors. 
The similar expansion of $C_v$ then
provides an alternative way of determining ${\bar\nu}$:
$\Delta C_v = C_v(T_1) - C_v(T_2) \propto N^{1/{\bar\nu}}$, 
leading to the estimation ${\bar\nu} \approx 2.0$ in Fig.~\ref{fig:mc}(d).

The divergence of the susceptibility when $T_c$ is approached
from above is described by the critical exponent $\gamma$:
$\chi \sim (T - T_c)^{-\gamma}$, which suggests
the finite-size scaling form 
\begin{equation}
\label{eq:chiscale}
\chi = N^{\gamma/{\bar \nu}} g\Bigl( (T{-}T_c)N^{1/{\bar\nu}}\Bigr)
\end{equation}
with the appropriate scaling function $g(x)$.
Combined with ${\bar\nu} \approx 2$ found above, the finite-size scaling 
form~(\ref{eq:chiscale}) yields the value 
$\gamma \approx 1$ and $T_c = 1.82(2)$ as shown in Fig.~\ref{fig:mc}(e).

Finally, on the basis of the above observation, the critical exponent
$\beta$ for the magnetization
is then determined from the scaling form
\begin{equation} \label{eq:m}
[\langle m \rangle]= N^{-\beta/\bar\nu} h\Bigl((T{-}T_c)N^{1/\bar\nu}\Bigr),
\end{equation}
which leads to $T_c = 1.82(1)$ and $\beta \approx 1/2$ [see Fig.~\ref{fig:mc}(f)]. 

We note that the obtained critical temperature appears to be higher 
by factor two than that in Ref.~\onlinecite{ref:Pekalski}.  
The standard mean-field approximation applied to this model, 
where the coordination number
is three, yields $T_{MF} = 3$~\cite{ref:Goldenfeld}. 
Considering the infinite range of
the interactions and the mean-field nature of the transition, 
we believe that our estimation $T_c = 1.82(2)$ (in units of $J/k_B$)
is more precise. In this respect, it is interesting to note 
that also in the $XY$ model on the small-world network a relatively large
value $T_c/T_{MF} \approx 0.9$ has been estimated~\cite{ref:XY}.

In summary, we have numerically studied the Ising model on the small-world
network constructed in the identical way as in Ref.~\onlinecite{ref:Pekalski}.
In contrast to Ref.~\onlinecite{ref:Pekalski}, we have obtained 
the standard mean-field critical exponents: $\alpha = 0, \beta = 1/2, \gamma = 1$,
and ${\bar \nu} = 2$ and confirmed that
the phase transition is of the mean-field nature, in agreement with other previous 
studies~\cite{ref:Barrat,ref:Gitt,ref:XY}.

\end{document}